\PassOptionsToPackage{table}{xcolor}
\documentclass[10pt,twocolumn]{IEEEtran}

\usepackage{amsmath,amssymb,amsfonts}
\usepackage{cite}
\usepackage{algorithm}
\usepackage{algorithmic}
\usepackage{graphicx}
\usepackage{textcomp}
\usepackage[table]{xcolor}

\makeatletter

\begin{document}
\title{Near-Field Localization with Physics-Compliant Electromagnetic Model: Algorithms and Model Mismatch Analysis
\thanks{A. M. Kuzminskiy, A. Elzanaty,  G. Gradoni, R. Tafazolli are with the 5GIC \& 6GIC, Institute for Communication Systems (ICS), University of Surrey, UK, 
(a.kuzminskiy,a.elzanaty,g.gradoni,r.tafazolli)@surrey.ac.uk.  F. Wang is with 
Huawei Technologies Co., Ltd., China,  fan.wang@huawei.com.}}
\author
{Alexandr M. Kuzminskiy, Ahmed Elzanaty, \emph{Senior Member, IEEE,} Gabriele Gradoni, \emph{Member, IEEE,} Fan Wang, Rahim Tafazolli, \emph{Fellow, IEEE}}

\maketitle
\thispagestyle{empty}
\pagestyle{empty}

\begin{abstract}
Accurate signal localization is critical for Internet of Things (IoT) applications, but precise propagation models are often unavailable due to uncontrollable factors. Simplified models, such as planar and spherical wavefront approximations, are widely used but can cause model mismatches that reduce accuracy. To address this, we propose an expected likelihood ratio (ELR) framework for model mismatch analysis and online (on-the-fly) model selection without requiring knowledge of the true propagation model. The framework leverages the scenario-independent distribution of the likelihood ratio of the actual covariance matrix, enabling the detection of mismatches and outliers by comparing given models to a predefined distribution. When an accurate electromagnetic (EM) model is unavailable, the robustness of the framework is analyzed using data generated from a precise EM model and simplified models within positioning algorithms. Validation in direct localization and reconfigurable intelligent surface-assisted scenarios demonstrates the framework’s ability to improve localization accuracy and reliably detect model mismatches in diverse IoT environments.
\end{abstract}


\section{Introduction}
The number of connected mobile devices and Internet of Things (IoT) is rapidly growing \cite{FarahsariSurveyIoT:22}. 
Radio localization supports various IoT emerging applications in 5G/6G communication networks \cite{iot2,6gold}  including 
autonomous driving \cite{int1,SomayahSignleLoc:24}, digital twins \cite{int2}, and augmented reality \cite{int3}.

Most localization algorithms rely on accurate models of the received signals as a function of the geometric channel parameters. In many practical scenarios, complete modelling of the received data may be difficult or impossible, for example, because of the unknown 
electromagnetic (EM) features or propagation scenarios such as antenna tilt and orientation at the transmitter \cite{surrey}. Many signal-processing papers ignore this situation and assume the same channel models for data generation and algorithm development, which is likely to ``yield overly optimistic and potentially misleading results" \cite{fried1}, \cite{fried2}. 

One way to overcome this difficulty could be a detailed EM propagation modeling in attempt to approach the true model (TM) for localization including modeling of tilt and orientation of transmit antennas, e.g., as in \cite{tilt}. EM-based models, such as those employing Green's function, offer a powerful approach to source localization by modelling wave propagation in complex environments. Green’s function methods enable the exact solution of the wave equation under various boundary conditions, facilitating more accurate modeling of electromagnetic fields in both near-field and far-field scenarios.  For instance, in \cite{CRLBHolographicEM:22,CRLBEM:23}, the authors apply Green's function to determine theoretical limits on positioning accuracy in holographic systems \cite{Elzanaty:Holographic:23}. 
However, approaching the TM by means of detailed EM modeling may be very difficult especially when it involves estimation of the transmitter parameters such that antenna tilt and orientation.

Another approach is using simplified localization models, 
e.g., near field (NF) and far field (FF) \cite{nf} models for point sources and different mutual coupling (MC) corrections \cite{fried3}, leading to the problem of model selection and mismatch analysis due to the mismatch between the data generation and  the presumed model for the localization algorithm \cite{surrey}.  In fact, localization approaches often employ FF models to estimate the positions of point sources where sources are sufficiently distant, enabling the assumption of planar wavefronts. However, with the introduction of massive MIMO, extremely large aperture arrays, and large reconfigurable intelligent surfaces, the users/sources may fall easily in the NF region, making the FF inaccurate \cite{Chen2024}.  NF models are a step forward as they account for the curvature of wavefronts and require more sophisticated processing techniques with usually 2D search over range and angle \cite{AnnaNFtracking:21,DardariFrancescoRIS:22}. Several works consider reducing the computational complexity by avoiding the 2D search over range and angle \cite{RootMUSIC:18,RinchiRIS:2022}. Such models can be enhanced by considering also the mutual coupling between the antennas \cite{fried2,fried3,MutuaCoulingEbafiZof:24}. 


One known way to analyze the localization performance under those mismatched models 
is using the misspecified Cramer-Rao lower bound (MCRB) as in \cite{mcrb,nf,Chen2024}, and others. Based on the full knowledge of the data generation  
TM and the simplified mismatched model (MM), MCRB allows to study degradation of the MM potential localization accuracy compared to the potential performance 
defined by means of the TM Cramer-Rao bound (CRB). 
The important difficulties of CRB based analysis are the following:
\begin{itemize}
\item The requirement of the full TM knowledge may be unrealistic in practical scenarios.
\item  CRB based analysis is not applicable for model selection and/or classification of the localization estimates for the given realization of the received signal, for example  
corresponding to different initialization of some optimization procedures such as the maximum likelihood (ML) algorithm. 
\end{itemize}

We propose an alternative framework to study model mismatch effects that does not require the TM knowledge and allows online model selection and classification of the 
localization estimates according to their correspondence to the unknown TM. The proposed framework considers the Expected Likelihood (EL) statistical approach 
developed for radar applications in \cite{el1,el3}, and others. 
The main EL idea is that under the usual Gaussian assumption of independent samples, the likelihood ratio (LR) of the actual (a priori unknown)
covariance matrix of the received signal is described by distribution that depends only on the problem dimension such as the number of samples and receive antennas, and 
it does not depend on the actual covariance matrix, i.e., it is scenario independent. This means that any matrix with an LR value inside this distribution is statistically as likely as the 
actual (true) covariance matrix. Any matrix with an LR value outside this distribution can be considered as statistically mismatched compared the unknown true matrix.

In the considered positioning application, this remarkable property allows to avoid the TM need-to-know requirement for online model mismatch analysis.  
It can be replaced with comparison of the LR value for the estimated covariance matrix 
with the predefined distribution of the actual covariance matrix.
If the obtained LR value belongs to the predefined distribution, then we can conclude that the MM and the corresponding localization 
estimate define the estimated covariance matrix that statistically is similar to the unknown actual covariance matrix. This allows us to classify the given MM and the corresponding
localization estimates as statistically relevant. Otherwise, they can be classified as outliers leading to a possibility to test other models, number of sources, initialization, etc., for the given 
realization of the received signal.

In this paper, we propose EL model mismatch analysis using the analytical EM model \cite{em1,em2,em3}  as the TM for direct and reconfigurable intelligent surface (RIS) \cite{ris} assisted localization.  
We analyze the EM propagation model features important for localization, select non-uniform array and RIS configurations and develop RIS profile optimization suitable for EM based localization. To concentrate on the possibility of avoiding the TM need-to-know requirement for online model mismatch analysis, we consider a fully digital receiving antenna array.\footnote{Note that our framework is also applicable to hybrid and holographic arrays with a reduced number of RF chains \cite{hybr,Haiyang:22, ElzanatyNFComm:24}.} Our contribution can be summarized as follows:
\begin{itemize}
\item  We propose a localization algorithm with EM true model that accounts for the mutual coupling between all the network elements that is valid for both the FF and NF scenarios.

\item We define the model mismatch analysis when mismatched models are considered for the localization algorithm that differs from the EM model considered for data generation.

\item We propose localization solutions including sparse arrays, RIS configurations, and impedance optimization for RIS-assisted localization for scenarios with limited line of sight (LOS). 

\item We conduct EL-based model mismatch analysis for direct and RIS-assisted localization and show some challenges in that method for EM modelling.
\end{itemize}
The rest of the paper is organized as follows. The signal and channel models and problem formulation are given in Section II. The basic EL formulation and signal-based offline EL model mismatch analysis are 
presented in Section III. 
The localization algorithms and online EL analysis are summarized in Section IV. The RIS-assisted localization with the base station (BS) array and RIS configurations selection and optimized RIS profile 
is addressed in Section V. 
The EM modeling based simulation results are given in Section VI. Section VII concludes the paper.

\section{Data model and problem formulation}
\begin{figure}[!t]
\centering
\includegraphics[width=70mm,height=50mm]{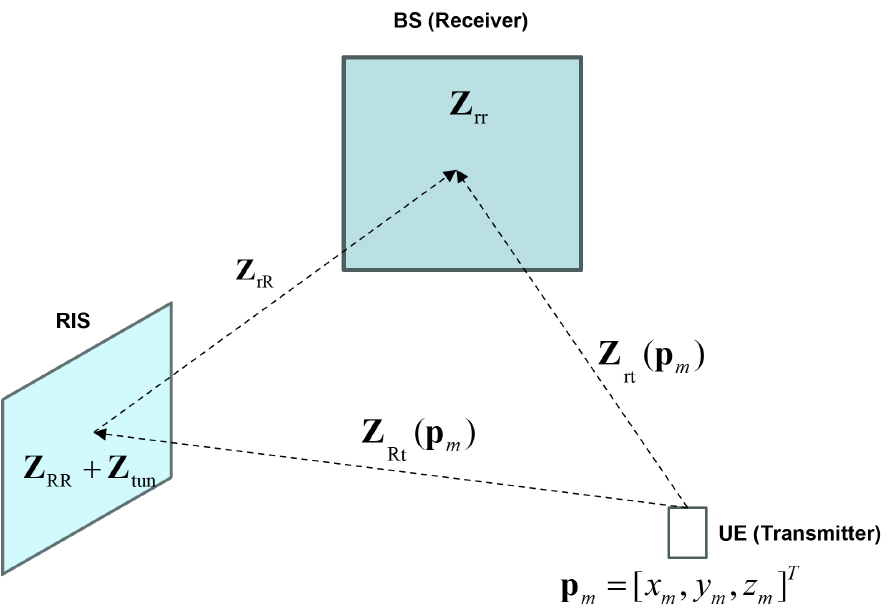} 
\caption{Considered scenarios: \textit{(i)}  UE-RIS and/or RIS/BS links are blocked for direct LOS scenario;  \textit{(ii)} UE-BS link is blocked for indirect LOS, RIS-assisted scenario.}
\label{fig:scenario}
\end{figure}

We consider an uplink system with BS equipped with $N=N_{\textrm{h}} N_{\textrm{v}}$ element planar antenna array
estimating the location of $M$ single-antenna sources where $N_{\text{h}}$ and $N_{\text{v}}$ are the numbers of the horizontal and vertical elements. We consider two scenarios in this paper: \textit{(i)} Direct LOS where the BS receives signals through direct source-to-BS link, \textit{(ii)} RIS-assisted where the BS receives signals through indirect source-to-RIS and RIS-to-BS paths with no direct link from the source to BS. For the channel model, we will first consider the general case with both direct LOS and non-direct LOS through RIS, because the special cases for scenarios \textit{(i)} and \textit{(ii)} can be derived from it, as in Fig.~\ref{fig:scenario}.

Initially, we consider uniform arrays for the BS and RIS with $d_{\text{h}}$, $d_{\text{v}}$ and  $d_{\text{hR}}$, $d_{\text{vR}}$ horizontal and vertical spacing for BS and RIS, respectively. 
Non-uniform array and RIS configurations are addressed in Section V.A.

The centre of the BS array is located at the origin, i.e., ${\bf r}_0 = [0,0,0]^{\text{T}}$ m, while the array coordinates ${\bf r}_n=[x_n,y_n,0]^{\text{T}}$, $n=1,...,N$ are as follows:
\begin{align} 
	x_n&=d_{\text{h}}\left(-\frac{N_{\text{h}}-1}{2}+\mbox{mod}(n-1,N_{\text{h}})\right), \label{eq:BS1}  \\
y_n&=d_{\text{v}}\left(\frac{N_{\text{v}}-1}{2}+\lfloor(n-1)/N_{\text{h}}\rfloor\right). \label{eq:BS2}  
\end{align}

A planar RIS consists of $N_{\text{R}}=N_{\text{hR}}N_{\text{vR}}$ elements,
where $N_{\text{hR}}$ and $N_{\text{vR}}$ are the number of the horizontal and vertical elements 
with the RIS origin location of ${\bf b}_0=[x_{\text{R}},y_{\text{R}},z_{\text{R}}]^T$ and $\phi$ degrees rotation in the $x,z$-plane. For the RIS element coordinates ${\bf b}_q=[x_q,y_q,z_q]^T$, $q=1,...,N_{\text{R}}$ we have
\begin{align} 
	x_{0q}&=d_{\text{hR}}\left(-\frac{N_{\text{hR}}-1}{2}+\mbox{mod}(q-1,N_{\text{hR}}\right),\label{eq:RIS1} \\
y_{0q}&=d_{\text{vR}}\left(\frac{N_{\text{vR}}-1}{2}+\lfloor(q-1)/N_{\text{vR}}\rfloor\right),\\
	y_q&=y_{\text{R}}+y_{0q},\\
x_q&=x_{\text{R}}+\mbox{cos}(\phi)x_{0q},\\
z_q&=z_{\text{R}}+\mbox{sin}(\phi)x_{0q}.\label{eq:RISend}
\end{align}

We consider $M$ sources with $m^{\text{th}}$ source location ${\bf p}_{m}=[x_{m},y_{m},z_{m}]^{\text{T}}, m=1,...,M$, 
where $(\cdot )^{\text{T}}$ is the transposition operation. 

\subsection{True Model for Received Signal}
The $N\times T$ matrix of the BS received signal ${\bf X}$ generated with the TM MIMO EM channel ${\bf H}_{\textrm{EM}}({\bf P})$ can be expressed as
 \begin{equation} 
	{\bf X}={\bf H}_{\textrm{EM}}({\bf P})\sqrt{\bf G}{\bf S}+{\bf U},
    \label{eq:X}
\end{equation}
where $T$ is the number of snapshots, ${\bf G}$ is the $M\times M$ diagonal matrix of signal power, 
${\bf S}$ is the $M\times T$ matrix of the unit power transmitted signals, ${\bf P}=[{\bf p}_{1},...,{\bf p}_{M}]$ is the $3\times M$ matrix of source locations, 
and ${\bf U}$ is the $N\times T$ matrix of additive thermal noise, where ${ {\bf u}_{t}}\sim {\cal{CN}}\{{\bf 0},\sigma^2  {\bf I}_N\}$, $\sigma^2$ is the noise power, and ${\bf I}_N$ is the 
$N\times N$ unit matrix.
As usually, for the EL and stochastic ML formulation, we assume ${\bf s}_t\sim {\cal{CN}}\{{\bf 0},{\bf I}_M\}$.\footnote{We use independent random 
finite alphabet symbols for simulations.}

Assuming that the array, RIS, and sources elements are vertical dipoles of $l$ length and $a$ radius, the true MIMO EM based model for the 
propagation channel ${\bf H}_{\textrm{EM}}({\bf P})$ is the $N\times M$ matrix defined according to the analytical EM model \cite{em1}, \cite{em2}:
\begin{align}
\label{eq:HEMTr}
	{\bf H}_{\text{EM}}({\bf P})&={\bf \Psi}_0^{-1}({\bf P}){\bf \Psi}_{\text{rt}}({\bf P}) \left[{\bf \Psi}_{\text{tt}}({\bf P})+{\bf Z}_{\text{t}}\right]^{-1},\\
    \label{eq:PSI0}
	{\bf \Psi}_0({\bf P})&={\bf I}_N\!+\!{\bf \Psi}_{\text{rr}}{\bf Z}_{\text{r}}^{-1}\!-\!{\bf \Psi}_{\text{rt}}({\bf P})\left[{\bf \Psi}_{\text{tt}}({\bf P})+{\bf Z}_{\text{t}}\right]^{-1}{\bf \Psi}_{\text{rt}}^{\text{T}}({\bf P}),\\
    \label{eq:PSIrr}
	{\bf \Psi}_{\text{rr}}&={\bf Z}_{\text{rr}}-{\bf Z}_{\text{rR}}({\bf Z}_{\text{RR}}+{\bf Z}_{\text{tun}})^{-1}{\bf Z}_{\text{rR}}^{\text{T}},\\
    \label{eq:PSIrt}
	{\bf \Psi}_{\text{rt}}({\bf P})&={\bf Z}_{\text{rt}}({\bf P})-{\bf Z}_{\text{rR}}({\bf Z}_{\text{RR}}+{\bf Z}_{\text{tun}})^{-1}{\bf Z}_{\text{Rt}}({\bf P}),
\end{align}
where  ${\bf Z}_{\text{t}}$ and ${\bf Z}_{\text{r}}$ are the $M\times M$ and $N\times N$ diagonal matrices of
the internal impedances of the transmit antennas and
the load impedances of the receive antennas, respectively, ${\bf Z}_{\text{rr}}$ and  ${\bf Z}_{\text{tt}}$ are 
the $N\times N$ and $M\times M$ matrices of mutual impedance between the BS array elements and the transmit antennas correspondingly, 
${\bf Z}_{\text{rt}}$ is the $N\times M$ matrix of mutual impedance between the BS and transmit antennas,  
${\bf Z}_{\text{RR}}$,  ${\bf Z}_{\text{rR}}$, and   ${\bf Z}_{\text{Rt}}$ are 
the $N_{\text{R}}\times N_{\text{R}}$, $N\times N_{\text{R}}$, and $N_{\text{R}}\times M$ matrices of mutual impedance between the RIS, the BS array and RIS, and 
the RIS and transmit antenna elements correspondingly,  and 
${\bf Z}_{\text{tun}}=\mbox{diag}(R_0+j{\bf f})$ is the $N_{\text{R}}\times N_{\text{R}}$ diagonal matrix, where $R_0>0$ and 
${\bf f}\in{\cal  R}^{1\times N_{\text{R}}}$ is the  vector of tunable impedance. 

For the direct source-to-BS link with no RIS,  \eqref{eq:PSIrr} and \eqref{eq:PSIrt} become
 \begin{equation} \label{eq:PSIrrdir}
	{\bf \Psi}_{\text{rr}}={\bf Z}_{\text{rr}},
\end{equation}
\begin{equation} \label{eq:PSIrtdir}
	{\bf \Psi}_{\text{rt}}({\bf P})={\bf Z}_{\text{rt}}({\bf P}).
\end{equation}
For the indirect signal-to-RIS and RIS-to-BS paths with no direct link to BS, \eqref{eq:PSIrt} becomes
\begin{equation} \label{eq:PSIrtRIS}
	{\bf \Psi}_{\text{rt}}({\bf P})=-{\bf Z}_{\text{rR}}({\bf Z}_{\text{RR}}+{\bf Z}_{\text{tun}})^{-1}{\bf Z}_{\text{Rt}}({\bf P}).
\end{equation}
It is worth pointing out that \eqref{eq:HEMTr} has been derived from first physics principle using the method of moments in electromagnetic, which extends the validity of
 the mathematical channel model beyond minimum scattering antennas, e.g., dipoles, and has been validated by full-wave simulations \cite{em4}.

All elements of the impedance matrices in \eqref{eq:HEMTr} - \eqref{eq:PSIrt} can be calculated as 
the mutual impedance $Z_{qp}$ between dipoles $q$ and $p$ located at $[x_q,y_q,z_q]^{\text{T}}$ and  $[x_p,y_p,z_p]^{\text{T}}$ correspondingly as defined in \cite{em3}:
\footnote{Following the array and RIS configurations defined in (1) and (2), we assume that all dipoles are aligned along the $y$-axis.}
\begin{align}
\!\!Z_{qp} &=\frac{{\eta {c_q}}}{{8\pi}}\!\!\!\sum\limits_{{s_0} = \{ - 1, + 1\} } {{s_0}} \exp \left( {j{s_0}k{h_q}} \right){I_{qp}}\left( {{\xi _p} = + {h_p};{s_0}} \right)\nonumber \\
&+ \frac{{\eta {c_q}}}{{8\pi }}\!\!\!\sum\limits_{{s_0} = \{ - 1, + 1\} } {{s_0}} \exp \left( {j{s_0}k{h_q}} \right){I_{qp}}\left( {{\xi _p} = - {h_p};{s_0}} \right) \nonumber\\
&- \frac{{\eta {c_{qp}}}}{{8\pi }}\!\!\!\sum\limits_{{s_0} = \{ - 1, + 1\} } {{s_0}} \exp \left( {j{s_0}k{h_q}} \right){I_{qp}}\left( {{\xi _p} = 0;{s_0}} \right),
\end{align}
where $\eta$ is the intrinsic impedance of free space, $c_q=1/\sin(kh_q)$, $c_{qp}=2\cos(kh_p)c_q$, 
$k=2\pi \lambda$, where $\lambda$ is the wavelength, $h_p$ and $h_q$ denote the half-length of the
$p$th and $q$th element,

\begin{equation} \label{eq:Iqp}
	{I_{qp}}\left( {{\xi _p};{s_0}} \right) = \mathcal{J}\left( { - {s_0},{\rho _{qp}},{\xi _p} - {y_{qp}}; - {h_q},0} \right) 
\end{equation}
\[
+ \mathcal{J}\left( { + {s_0},{\rho _{qp}},{\xi _p} - {y_{qp}};0, + {h_q}} \right), 
\]
\begin{equation}
	\mathcal{J}\left( {{s_0},{d_0},{y_0};L,U} \right) = {s_0}\exp \left( { - jk{s_0}{y_0}} \right){E_1}\left( {jk{L_0}} \right)
\end{equation}
\[
- {s_0}\exp \left( { - jk{s_0}{y_0}} \right){E_1}\left( {jk{U_0}} \right), 
\]
\begin{equation}
	{L_0} = \sqrt {d_0^2 + {{\left( {L - {y_0}} \right)}^2}} + {s_0}\left( {L - {y_0}} \right) 
\end{equation}
\begin{equation}
 {U_0} = \sqrt {d_0^2 + {{\left( {U - {y_0}} \right)}^2}} + {s_0}\left( {U - {y_0}} \right), 
\end{equation}
\begin{equation}{E_1}(c) = \int_c^\infty {\frac{{\exp ( - u)}}{u}} du,
\end{equation}
where $\rho_{qp}^2=(x_q-x_p)^2+(z_q-z_p)^2$ if $p\ne q$ and  $\rho_{qp}^2=a^2$ if $p=q$, respectively, 
$y_{qp}=y_q-y_p$, and ${E_1}(c) $ is the exponential integral \cite{exp}.

If $\rho_{qp}=0$, e.g., the elements are in a collinear formation, $E_1(c)$ is not defined for $c=0$. Then, 
${I_{qp}}\left( {{\xi _p};{s_0}} \right) $ in \eqref{eq:Iqp} can be obtained by means of numerical integration:
\begin{multline}
    {I_{qp}}\left( {{\xi _p};{s_0}} \right) = \int_{-h_q}^0\frac{ \exp (-jk |y+y_{qp}-y_p|)}{ |y+y_{qp}-y_p|} \exp (js_0 ky) dy \\
    +\int_0^{h_q} \frac{ \exp (-jk |y+y_{qp}-y_p|)}{ |y+y_{qp}-y_p|} \exp (-js_0 ky) dy. 
\end{multline}

In all our analysis and simulation, we consider that the received signal is according to the EM-model in \eqref{eq:X}. However, for the localization algorithm (e.g., the array manifold and the likelihood function), we can consider either the true EM model or mismatched models when the true model is not known or is too complex to be employed.
\subsection{Mismatched Models}
The spherical wavefront NF and plane wave FF mismatched models are defined for single source channels because we do not consider mutual coupling between transmit antennas for MM. We omit a source index for simplicity and define the $n^{\text{th}}$ element of the NF and FF channels as follows.
\subsubsection{NF Mismatched Model}
The channel from the source transmit antenna to the $n^\text{th}$ BS antenna can be written considering the NF spherical wavefront model as 
\begin{equation}
	h_n^{\textrm{NF}}({\bf p},{\bf r}_n)=\frac{\lambda}{4\pi||{\bf p}-{\bf r}_n||}e^{-j\xi} e^{-j\frac{2\pi}{\lambda}||{\bf p}-{\bf r}_n||},
\end{equation}
where $\lambda$ is the wavelength and $\xi$ is the initial phase for the channel. We can notice that the exact distance between the source and $n^\text{th}$ BS antenna, i.e.,  $||{\bf p}-{\bf r}_n||$, is accounted for in both the channel gain and the induced phase shift. This captures the wavefront's spherical nature and the spatial non-stationarity for the amplitude across various antennas at the BS array. The channel can be rewritten as
\begin{align}
	h_n^{\textrm{NF}}({\bf p},{\bf r}_n)&={\alpha}_{n}({\bf p})\,{d}_{n}({\bf p})\,D({\bf p}),\\
	\alpha_n({\bf p})&=\alpha \frac{||{\bf p}||}{||{\bf p}-{\bf r}_n||},\\
	d_n({\bf p})&=e^{-j\frac{2\pi}{\lambda}(||{\bf p}-{\bf r}_n||-||{\bf p}||)},
\end{align}
where $\alpha=\frac{\lambda}{4\pi||{\bf p}||}e^{-j\xi}$ and $D({\bf p})=e^{-j\frac{2\pi}{\lambda}||{\bf p}||}$. The channel between the source and the $q^{\text{th}}$ RIS element can be expressed similarly.  
\subsubsection{FF Mismatched Model}
The channel from the source transmit antenna to the $n^\text{th}$ BS antenna can be written considering the FF plane wavefront model as \cite{ff}
\begin{equation}
	 h_n^{\textrm{FF}}({\bf p},{\bf r}_n)=\alpha D({\bf p}) e^{-j\varphi_n}, 
\end{equation}
where $\varphi_n\!=\!{\bf v}^T(\phi,\psi){\bf r}_n$, $\phi\!=\!\mbox{tan}^{-1}(-x_s/z_s)$, $\psi=\mbox{tan}^{-1}(-y_s/\sqrt{x_s^2+z_s^2})$,  and 
${\bf v}=[\mbox{cos}(\psi)\mbox{sin}(\phi),\mbox{sin}(\psi),\mbox{cos}(\psi)\mbox{cos}(\phi)]^T$. For the FF model, the first Taylor approximation is considered for the difference in the distances between the source-BS array centre and source-BS $n^\text{th}$ antenna, rather than the exact distance. Also, we can notice that the channel amplitude is constant across various BS antenna elements and does not depend on the antenna index $n$. 
Similarly, the source-RIS channel can be expressed after adjusting for the origin and rotation angle of the RIS.
\subsubsection{Mismatched Models with Mutual Coupling Correction}
 We also consider an approximate linear MC correction matrix, ${\bf C}_{\textrm{MC}}$, that does not depend on the source location, where the channel model can be modified to account for the MC between the BS array antenna elements using the previously expressed mismatched models as
 \begin{equation} \label{eq:hMMMC}
	{\bf h}^{\textrm{MMMC}}({\bf p})={\bf C}_{\textrm{MC}}\,{\bf h}^{\text{MM}}({\bf p}),
\end{equation}
for NF and FF used as MM. For the source-BS link, the MC correction matrix can be written as \cite{fried3}
\begin{equation} \label{eq:Cmc}
	{\bf C}_{\textrm{MC}}={\bf Z}_{\text{r}} ({\bf Z}_{\text{r}}+{\bf Z}_{\text{rr}})^{-1}.
\end{equation}
On the other hand, for RIS-assisted localization, we have 
\begin{equation} \label{eq:CmcRIS}
	{\bf C}_{\textrm{MC}}=-{\bf Z}_{\text{r}} ({\bf Z}_{\text{r}}+{\bf Z}_{\text{rr}})^{-1}{\bf Z}_{\text{rR}}({\bf Z}_{\text{RR}}+{\bf Z}_{\text{tun}})^{-1}.
\end{equation}
Note that all the elements of the impedance matrices in \eqref{eq:Cmc}, \eqref{eq:CmcRIS} depend on the BS array and RIS dipole positions, which are assumed to be known. 
It is worth emphasizing that compared to the TM, the considered MMs do not use the  
$ {\bf Z}_{\text{rt}}$ and $ {\bf Z}_{\text{tt}}$ impedance matrices that depend on the potentially unknown tilt and orientation of the transmit antennas. Therefore, those mismatched models do not account for the mutual coupling between the transmit and receive antennas. 

\medskip

Our problem is to analyze the applicability of the mismatch models NF, FF, NFMC, and FFMC assuming the analytical EM model 
in equation \eqref{eq:X} as the TM for data generation. 
To do that, we propose the EL as a metric to detect whether  there is a model mismatch or non accurate model parameter, i.e., position esitmate. 
This can allow us to detect outliers in real-time (online outlier detection) without the TM knowledge requirement.
Furthermore, for the given EM data generation model, we investigate using different BS array and RIS configurations and tunable impedance 
optimization algorithms for RIS-assisted localization.

\section{EL based model mismatch analysis}
First, we summarize the basic EL formulation to emphasize the main EL property of scenario independence of the actual covariance matrix LR distribution 
that allows to avoid the TM knowledge requirement in model mismatch analysis. Then, we introduce signal based EL model mismatch analysis 
if a possibility to generate the TM signal samples using any physical or program tools may be available.
  
\subsection{Basic EL formulation}
The likelihood ratio (likelihood function divided by its maximum value) for the $N\times T$ matrix of Gaussian received signal $\bf X$ and some non-singular model of its supposed covariance matrix 
${\bf R}_{\textrm{M}}$ (either the true or a mismatched model), for the undersampled case of $N>T$, can be presented as in \cite{el3}\footnote{We consider the undersampled case because it corresponds 
to the typical scenario for future networks with a high number of BS antennas and a low number of signal samples. The oversampled LR formulation can be found, e.g., in \cite{el1}. It would be suitable for hybrid or holographic versions of localization networks with a lower number of RF chains.}
\begin{equation} \label{eq:LRX}
	\mbox{LR}({\bf X}|{\bf R}_{\textrm{M}})=\left(\frac{\mbox{det}({\bf X}^*{\bf R}_{\textrm{M}}^{-1}{\bf X}/N)\mbox{exp}(T)}
       { \mbox{exp}(\mbox{tr}[{\bf X}^*{\bf R}_{\textrm{M}}^{-1}{\bf X}/N])}\right)^T,
\end{equation}
where $(\cdot)^*$ is the transposition conjugate operation, $\mbox{det}({\bf A})$, and  $\mbox{tr}({\bf A})$ are determinant and trace of matrix $\bf A$. 

The interesting feature of the LR ratio that allows us to analyze model mismatch is that it has the following property.  If the preassumed model for the covariance matrix of the received signal matches the actual covariance matrix of the received signal, i.e., no model mismatch, the probability density function (p.d.f.) of the LR will rely only on the problem dimension, i.e., $N$ and $T$, and not on the channel model, i.e., the model covariance matrix. 

In our case, it means that if ${\bf R}_0={\bf R}_\textrm{TM}$, where
\begin{equation}
	{\bf R}_0={\bf H}_{\textrm{EM}}({\bf P}){\bf G}{\bf H}_{\textrm{EM}}^*({\bf P})+\sigma^2{\bf I}_N
\end{equation}
is covariance matrix of the actual received signal computed from \eqref{eq:X}, then from the scenario-independence property of the LR (p.d.f.) for the actual covariance matrix
\begin{equation} \label{eq:pdf}
	\mbox{p.d.f.}\left[ \mbox{LR}({\bf X}|{\bf R}_0\right]=f(N,T)\ne f({\bf R}_0).
\end{equation}

Considering that the EL analysis uses the comparison between the received signal-based LR and the predefined LR, to simplify calculations, we use the same monotonic function on both of them $\mbox{LR}^{1/T}$ and keep the previous notation of LR. In particular, the (p.d.f.) of $\alpha_0\triangleq \text{LR}^{1/T}$  can be written in closed form from \cite{el3} as
\begin{equation}
	\!\!\!w(\alpha_{0})\!=\!C(T,N)\alpha_{0}^{N-T} G_{T,T}^{T,0}\!\left(\alpha_0|_{0,1,...,T-1}^{\frac{T^2-1}{T},\frac{T^2-2}{T},...,\frac{T^2-T}{T}}\right),
    \label{eq:alpha0PDF}
\end{equation}
where
\[
	C(T,N)=(2\pi)^{\frac{1}{2}(T-1)}T^{\frac{1}{2}-TN}\frac{\Gamma(TN)}{\prod_{j=1}^T \Gamma(N-j+1)},
\]
 $\Gamma(x)$ represents the Gamma function \cite{formula}, and $G_{c,d}^{a,b}(\cdot)$ is Meijer’s G-function \cite{formula}. This confirms that the LR p.d.f when the true model is considered can be computed solely by knowing $N$ and $T$. Another way to obtain the p.d.f. in \eqref{eq:alpha0PDF} is by relying on the model independence property for it by the generation of a high number of the unit power AWGN 
$N\times T$ matrices $\bf E$ and calculation of histogram of the corresponding LR samples 
\begin{equation} \label{eq:LRE}
	{\mbox{LR}}({\bf E}|{\bf I}_N)=\frac{\mbox{det}({\bf E}^*{\bf E}/N)\mbox{exp}{(T)}}
       { \mbox{exp}(\mbox{tr}[{\bf E}^*{\bf E}/N])}
\end{equation}
considering that ${\bf R}_0={\bf I}_N$ in the AWGN case. Fig. 2 presents the ${\mbox{LR}}$ distributions for the actual covariance matrix for $N=64$ and different $T$ estimated over 10000 realizations according to \eqref{eq:LRE}.
\begin{figure}[!t]
\centering
\includegraphics[width=70mm,height=50mm]{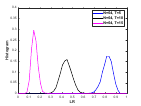} 
\caption{Pre-known ${\mbox{LR}}$ distributions for the actual covariance matrix for $N=64$ and different $T$.}
\label{fig:lr_distr_3}
\end{figure}

From the properties of the LR, our model mismatch analysis is based on: 1) calculation of  ${\bf R}_{\textrm{M}}$ for some array manifold models, not necessarily the correct one,  and their estimated parameters (e.g., estimated source location); 2) calculation of the corresponding LR value for the given signal realization according to equation \eqref{eq:LRX}; and 3) compare it with the predefined scenario-independent $\mbox{p.d.f.}\left[ \mbox{LR}({\bf X}|{\bf R}_{\textrm{TM}})\right]$ for the given
problem dimension parameters $N$ and $T$ computed from either \eqref{eq:alpha0PDF} or \eqref{eq:LRE}.

\subsection{Signal based EL model mismatch analysis}
A typical way to specify applicability areas for mismatched models is using Fraunhofer distance, e.g.,  \cite{nf}, although localization efficiency of different MMs may also 
depend on other scenario parameters such as signal-to-noise ratio (SNR), signal bandwidth, sampling support, etc.     
The EL approach allows to judge the statistical relevance of the given model in different scenarios.  We specify a signal based EL procedure assuming a possibility to generate the TM signal samples 
using a physical signal source,  the analytical EM model such as the one specified in Section II, 
or any other model for signal generation, e.g., based on a Numerical Electromagnetic Code (NEC) as suggested in \cite{fried2}.
It is worth emphasizing that we do not assume the TM knowledge, which potentially could be directly used for localization. 
For a signal based EL model mismatch analysis, we only assume a possibility of the TM signal generation.

For such EL analysis, 
all we need is to find a possibility to get the LR values in the predefined LR area for the actual covariance matrix for the given model.  
To do that, we need to generate signal samples and use the given MM with the known source location and  
scenario parameters such as the number of samples, signal and noise power to find the corresponding LR value. 
Then, we need to compare the LR value with the predefined distribution $\mbox{LR}({\bf X}|{\bf R}_0)$ for the given $N$ and $T$.

We propose the following signal based metric for EL model mismatch analysis
\begin{equation} \label{eq:MMsig}
	\hat{\gamma}_{\textrm{MM}}({\bf p}_s)=\frac{\frac{1}{K}\sum_{k=1}^K \mbox{LR}[{\bf X}_k|{\bf R}_{\textrm{MM}}({\bf p})]}
	{\mathbb{E}\{\mbox{LR}({\bf X}|{\bf R}_0)\}(N,T)},
\end{equation}  
where
${\bf X}_k$ is the $k^{\text{th}}$ TM signal realization, $K$ is the number of signal realizations, $\mathbb{E}{}$ is the expectation operator,
\begin{equation}
	{\bf R}_{\textrm{MM}}({\bf p})=g{\bf h}_{\textrm{MM}}({\bf p}){\bf h}_{\textrm{MM}}^*({\bf p})+\sigma^2{\bf I}_N
\end{equation}  
is the covariance matrix for the given MM with scenario parameters $g$ and $\sigma^2$.

It is clear that $\hat{\gamma}_{\textrm{TM}}({\bf p}) \rightarrow 1$ for $K\rightarrow \infty$. 
Thus, in the source location areas with $\hat{\gamma}_{\textrm{MM}}({\bf p}) \approx 1$ or $\hat{\gamma}_{\textrm{MM}}({\bf p}) \ll 1$ one can expect low or 
significant model mismatch effect compared to the TM, respectively. 

\section{Localization algorithm and online EL analysis}
Following our main assumption of the unknown TM for localization, we introduce a localization algorithm based on some MM that allows online classification 
of the given MM and corresponding localization estimates as statistically relevant or outliers. In the last case, other MMs and/or algorithm parameters such as 
initialization could be tested for the given received signal realization in attempt to obtain the statistically relevant results.
 
 We propose a stochastic ML optimization to estimate the position of the sources for the localization problem and to estimate the parameters of the model required for the EL. In this regard,
 the likelihood function for any non-singular parametric covariance matrix model ${\bf R}_{\textrm{MM}}$ is derived as
\begin{equation}
	{\cal{L}}({\bf X},{\bf R}_{\textrm{MM}})=\left( \frac{1}{\pi^N \mbox{det}({\bf R}_{\textrm{MM}})} 
	\mbox{exp}(-\mbox{tr}[{\bf R}_{\textrm{MM}}^{-1}\hat{\bf R}])  \right)^T,
\end{equation}  
where 
\begin{equation}
	\hat{\bf R}=\frac{1}{T}{\bf X}{\bf X}^*
\end{equation}  
is the sample covariance matrix of the received signal \cite{lf}, and
\begin{equation}
	{\bf R}_{\textrm{MM}}={\bf H}_{\textrm{MM}}({\bf P})\hat{\bf G}{\bf H}_{\textrm{MM}}^*({\bf P})+\sigma^2{\bf I}_N
\end{equation}
is the covariance matrix model for the given location $\bf P$ and MM channel matrix ${\bf H}_{\textrm{MM}}$ that consists of the corresponding channel vectors defined in \eqref{eq:hMMMC}. Assuming the known $\sigma^2$, the estimated source power can be estimated as
\begin{align} \label{eq:G}
	\hat{\bf G}&=\max [\mbox{diag}({\bf B}),0],\\
	{\bf B}&={\bf H}_{\textrm{MM}}^{\dag}({\bf P})(\hat{\bf R}-\sigma^2 {\bf I}_N){\bf H}_{\textrm{MM}}^{{\dag}*}({\bf P}),
\end{align}  
where $\mbox{diag}({\bf B})$ is the diagonal matrix of the diagonal elements of matrix $\bf B$, and 
$(\cdot)^{\dag}$ is the pseudoinverse operation \cite{power}.

Then, the stochastic ML localization algorithm can be formulated as\footnote{Another ML formulation is given, e.g., in \cite{ap}, which allows computationally simpler implementation by means of alternating projection algorithm.}
\begin{equation} \label{eq:PMM}
	\hat{\bf P}^{\textrm{MM}}=\mbox{arg}\min_{{\bf P}}
		 \left( \mbox{log} (\mbox{det}[{\bf R}_{\textrm{MM}}({\bf P})]) +\mbox{tr}[{\bf R}_{\textrm{MM}}^{-1}({\bf P})\hat{\bf R}]\right).
\end{equation}

The optimization procedure in \eqref{eq:PMM} needs initialization such that the MUSIC algorithm \cite{music}, which can be summarized for the given MM as  finding $\hat M$ peaks of the MUSIC spectrum:
 \begin{equation} \label{eq:music}
	f({\bf p})=\frac{1}{{\bf h}_{\textrm{MM}}^*({\bf p}){\bf V}_{\text{n}}{\bf V}_{\text{n}}^*{\bf h}_{\textrm{MM}}({\bf p})},
\end{equation}  
 where $\hat M$ is the estimated number of signals, e.g. using some of the information theoretic criteria such as minimum description length (MDL) \cite{mdl}, 
${\bf V}_{\text{n}}$ is the $N\times (N-\hat{M})$ matrix of the eigenvectors corresponding to the 
$N-\hat{M}$ smallest eigenvalues of the sample covariance matrix $\hat{\bf R}$.

Any array manifold, e.g.,  specified in Section II, and any local optimization algorithm with some, e.g., MUSIC, initialization can be used for maximization in \eqref{eq:PMM} without any guarantee of finding the 
global ML solution. The EL analysis of the given received signal realization allows the detection of the optimization outliers and opens the possibility of online initialization correction and/or model selection. It can be summarized in Algorith~\ref{alg:model_location_classification}.
\begin{algorithm}[t!]
\caption{Model and Location Estimates Classification}
\label{alg:model_location_classification}
\begin{algorithmic}[1]
\REQUIRE $N$, $T$, ${\bf X}$, ${\bf H}_{\textrm{MM}}$
\STATE Define the p.d.f. of the LR for the actual covariance matrix for the given $N$ and $T$ as described in Section III.A.
\STATE Estimate the number of sources in the received signal realization, e.g., using the information-theoretic criteria \cite{mdl}, and find location initialization, e.g., using the MUSIC algorithm \eqref{eq:music} defined over some search grid in the localization area of interest.
\STATE Find the localization estimate $\hat{\bf P}^{\textrm{MM}}$ for the given MM using the ML localization algorithm \eqref{eq:PMM}.
\STATE Calculate the LR value for the given MM and estimated location as
\begin{equation}
     \label{eq:LRMM}
     \!\!\!\mbox{LR}[{\bf X}|\hat{\bf R}_{\textrm{MM}}(\hat{\bf P}^{\text{MM}})]
\!=\! \left(\!\frac{\mbox{det}({\bf X}^*\hat{\bf R}_{\textrm{MM}}^{-1}(\hat{\bf P}^{\textrm{MM}}){\bf X}/N)\mbox{exp}(T)}
{\mbox{exp}(\mbox{tr}[{\bf X}^*\hat{\bf R}_{\textrm{MM}}^{-1}(\hat{\bf P}^{\textrm{MM}}){\bf X}/N])}\!\right)^T,
\end{equation}
where 
\[
\hat{\bf R}_{\textrm{MM}}(\hat{\bf P}^{\textrm{MM}}) = {\bf H}_{\textrm{MM}}(\hat{\bf P}^{\textrm{MM}})
\hat{\bf G}{\bf H}_{\textrm{MM}}^*(\hat{\bf P}^{\textrm{MM}}) + \sigma^2{\bf I}_N.
\]
\IF{
\begin{equation}
\label{eq:beta1}
\mbox{LR}[{\bf X}|\hat{\bf R}_{\textrm{MM}}(\hat{\bf P}^{\textrm{MM}})] > \beta(N,T)
\end{equation}
}
\STATE Classify the location estimate and the corresponding MM as statistically reliable, where $\beta(N,T)$ is the threshold for some probability $p_{\beta} \ll 1$ such that 
\[
\mbox{prob}\{\mbox{LR}({\bf X}|{\bf R}_0) < \beta(N,T)\} = p_{\beta},
\]
which can be determined based on the predefined $\mbox{p.d.f.}\left[\mbox{LR}({\bf X}|{\bf R}_0)\right]$.
\ELSIF{
\begin{equation}
\label{eq:beta2}
\mbox{LR}[{\bf X}|\hat{\bf R}_{MM}(\hat{\bf P}^{MM})] \ll \beta(N,T)
\end{equation}
}
\STATE Classify the location estimate and/or the corresponding MM as an outlier.
\ENDIF
\ENSURE The model and location estimates classified.
\end{algorithmic}
\end{algorithm}
%
Once the algorithm defines the localization estimation as an outlier, another model and/or initialization can be tested for the given signal realization in an attempt to remove the outlier. 
One possibility of such advanced model options could be using a mutual impedance between the transmit and receive antennas for different transmit antenna tilts and orientations 
instead of the using a point source models ${\bf h}^{\text{MM}}({\bf p})$ in \eqref{eq:hMMMC}.

It is important to note that even if the EM TM could be known, then using it in the localization algorithms \eqref{eq:PMM}, \eqref{eq:music} could require much higher computational complexity compared to MMs because it would involve computation of the mutual transmitter/array impedance depending on the transmitter location instead 
of using computationally simple geometrical NF or FF models, possibly with MC correction. This situation emphasizes the importance of model mismatch analysis.

\section{Model mismatch analysis for RIS-assisted localization}
If the LOS channel between the transmitter and BS array is obstructed, then localization ability could be restored via RIS-assisted localization. When considering the accurate analytical EM model for RIS, as presented in Section II, it becomes evident that positioning accuracy heavily depends on: \textit{(i)} the configuration of both BS array and RIS; \textit{(ii)} the RIS profile in terms of tunable impedance for the elements. In this section, we discuss and propose various methods to improve localization performance by addressing these factors. More specifically, in Section V.A, we examine the EM model features that may significantly impact localization performance and propose using non-uniform planar array configurations to address these features.

Notably, the analysis presented for EL in Section IV is directly applicable in the case of RIS-assisted localization, as the predefined likelihood ratio distribution does not rely on the RIS configuration or the tunable RIS profile. Nevertheless, the localization accuracy critically depends on the RIS profile. Therefore, in Section V.B, we develop an algorithm for the tunable impedance of RIS elements, assuming the specified electromagnetic transmission model for signal generation and considering that only mismatched models may be available for optimisation.

\subsection{Antenna configuration for RIS-assisted localization}
The accurate EM propagation modeling for RIS-assisted environments highlights scenarios where positioning becomes inherently challenging. One potential reason is that channels from different locations may exhibit similarities, making it difficult to distinguish between them. The most problematic case arises when the channels from different locations are nearly identical, differing only by a complex scalar. In such cases, the corresponding covariance matrices for these locations may also appear similar, scaled by the estimated signal power. This results in location uncertainty, as the associated LR values can align with the predefined LR distribution of the unknown actual covariance matrix due to their statistical similarity.

Let us illustrate this situation for $f_{\text{c}}=28$ GHz carrier frequency assuming vertical dipoles for BS array and RIS with $l=\lambda/2$, $a=\lambda/500$, and 
${\bf Z}_{\text{r}}=50{\bf I}_N$. Similar dipoles are used for single antenna transmitters with  ${\bf Z}_{\text{t}}=50{\bf I}_M$. We use $T=10$ 
independent random QPSK symbols as a transmitted signal realization. Sources are located in x,z-plane for $y=-h$ m below the BS array. For each source, we estimate three parameters $\hat{x}$, $\hat{z}$, and ${\hat{g}}$.
We consider uniform planar BS array of $N_{\text{h}}=N_{\text{v}}=8$, $N=64$, and  uniform planner RIS with $N_{\text{hR}}=N_{\text{vR}}=10$, $N_{\text{R}}=100$, 
$[x_{\text{R}},y_{\text{R}},z_{\text{R}}]=[1,0,1]$ m, $\phi=\pi/2$, spaced by $\lambda/2$ 
for the EM propagation channels modeled by equations \eqref{eq:HEMTr} - \eqref{eq:PSIrr}, \eqref{eq:PSIrtRIS} with $h=1$m, $R_0=0.2$, and $\sigma^2=-120$ dBm.

To evaluate a channel similarity for different locations, we define a metric representing the normalized mean square error (MSE) between the actual source location channel 
${\bf h}_{\text{EM}}({\bf p}_0)$ and its approximation 
by the other location channel ${\bf h}_{\text{EM}}({\bf p})$ up to the optimal in the least squares (LS) sense complex scalar $q({\bf p},{\bf p}_0)$:
\begin{equation} \label{eq:Q}
	Q({\bf p},{\bf p}_0)=\frac{||q({\bf p},{\bf p}_0){\bf h}_{\text{EM}}({\bf p})-{\bf h}_{\text{EM}}({\bf p}_0)||^2}{||{\bf h}_{\text{EM}}({\bf p}_0)||^2},
\end{equation}
where
\begin{equation}
	q({\bf p},{\bf p}_0)=\frac{{\bf h}^*_{\text{EM}}({\bf p}_0){\bf h}_{\text{EM}}({\bf p})}{||{\bf h}_{\text{EM}}({\bf p})||^2}.
\end{equation}
 
An example  of Q-metric \eqref{eq:Q} and the corresponding MUSIC pattern for NFMC as MM in \eqref{eq:music} for 10 dBm signal power and source located at ${\bf  p}_0=[-1.51,-1,3.1]^{\text{T}}$ m 
are plotted in Fig. 3 for a random tunable RIS profile. 
\begin{figure}[!t]
\centering
\includegraphics[width=99mm,height=37mm,clip]{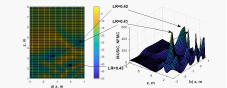} 
\caption{Uniform BS array and RIS for $10$ dBm signal power: a) Q-metric in dB for ${\bf p}_0=[-1.51,-1,3.1]^{\text{T}}$ m, b) MUSIC spectrum.}
\label{fig:Q}
\end{figure}
One can see in Fig.~\ref{fig:Q}a that there are many locations with channels that are very close in the Q-metric \eqref{eq:Q} sense to the channel for the actual location leading to the corresponding MUSIC peaks 
in Fig. 3b, although, only $\hat{M}=1$ signal can be detected using the MDL criterion for the estimated $T\times T$ covariance matrix ${\bf X}^*{\bf X}/N$. Furthermore, the LR values corresponding to such peaks belong to the corresponding predefined LR distribution in Fig. 2 for $N=64$ and $T=10$ with typical values between $0.35$ to $0.55$. Clearly, 
such a propagation environment is  challenging for signal localization. 

To relax this difficulty, we consider higher aperture array and RIS non-uniform configurations keeping the same $N$ and $N_{\text{R}}$ number of elements and minimum element spacing of 
$\lambda/2$ as for the uniform configurations, and using the MC correction defined in \eqref{eq:CmcRIS}. The largest aperture can be achieved for the minimum redundancy non-uniform configurations, e.g., \cite{nua}. For the considered $N_{\text{h}}=N_{\text{v}}=8$  and
 $N_{\text{hR}}=N_{\text{vR}}=10$ configurations 
the corresponding minimum redundancy element spacing in terms of the minimum $\lambda/2$ element distance  can be found from \cite{nua} as: $[1,3,6,6,2,3,2]$ and $[1,2,3,7,7,7,4,4,1]$ leading to $24\times24$ and 
$37\times 37$ non-uniform configurations for the BS array and RIS, respectively. 
These configurations can be implemented according to equations \eqref{eq:BS1} - \eqref{eq:BS2} and \eqref{eq:RIS1} - \eqref{eq:RISend} keeping dipoles only in the positions corresponding to the indicated patterns.

Similar to Fig. 3 plots for the indicated non-uniform configurations are shown in Fig. 4 for the same signal position and tunable RIS profile.
\begin{figure}[!t]
\centering
\includegraphics[width=99mm,height=37mm]{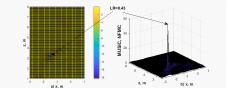} 
\caption{Non-uniform BS array and RIS: a) Q-metric in dB for ${\bf  p}_0=[-1.51,-1,3.1]^{\text{T}}$ m, b) MUSIC pattern for 10 dBm signal power.}
\label{fig:Qmetricnonuniform}
\end{figure}
One can see in Fig.~\ref{fig:Qmetricnonuniform} emonstrates that employing higher aperture configurations with non-uniform MC correction as per \eqref{eq:CmcRIS} significantly reduces the number of locations where channels are similar, up to a complex scalar, to the actual location channel. This improvement leads to a markedly enhanced MUSIC signal peak pattern, with the corresponding LR values conforming to the predefined LR distribution. The primary reason is that increasing the aperture of the BS array and RIS extends the Fraunhofer distance, positioning the source in the near-field. This shift provides more degrees of freedom and enhances the channel's uniqueness.

While these configurations cannot guarantee such channel properties for every signal location and RIS profile, the example illustrates their ability to significantly expand the signal location areas that enable reliable localization. These configurations are employed in the RIS-related simulations in Section VI.C.

\subsection{Tunable RIS profile optimization}
\label{subsec:tunableRIS}
In this subsection, we propose an algorithm for RIS-assisted localization with optimizing impedance for the tunable RIS elements. 
We propose a two-stage localization procedure, illustrated in Fig. 5. For the first $T_1$ samples, we apply a random RIS profile $\tilde{\bf f}$, 
estimate the number of sources $\tilde{M}$ 
by means of the MDL criterion, obtain the localization estimate $\tilde{\bf P}(\tilde{\bf f})$  
using the MUSIC algorithm \eqref{eq:music},  and optimize the RIS profile $\hat{\bf f}$ for the fixed $R_0$ in the tunable impedance matrix ${\bf Z}_{\text{tun}}$ in \eqref{eq:HEMTr} - \eqref{eq:PSIrr}, \eqref{eq:PSIrtRIS}. 
Then, we apply $\hat{\bf f}$, receive the next $T_2$ samples and obtain the final localization estimate 
 $\hat{\bf P}$ by means of the ML algorithm \eqref{eq:PMM}.
\begin{figure}[!htb]
\centering
\includegraphics[width=50mm,height=15mm]{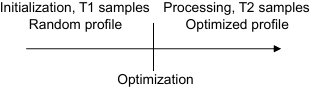} 
\caption{RIS optimization protocol.}
\label{fig:ris_opt_1}
\end{figure}

For the best localization performance, CRB optimization for the initial localization estimates should be used, but the TM knowledge would be needed for that. 
For the unknown TM and the given mismatched localization model, e.g., NFMC in \eqref{eq:hMMMC}, we approximate the CRB optimization as
\begin{equation} \label{eq:opt}
	\hat{\bf f}=\mbox{arg} \min_{{\bf f}\in R^{1\times N_{\text{R}}}}\mbox{CRB}_{\textrm{NFMC}}[\tilde{\bf P}(\tilde{\bf f}),{\bf f}],
\end{equation}  
where
\begin{equation}
	\mbox{CRB}_{\textrm{NFMC}}({\bf q},{\bf f})=\sqrt{\mbox{tr}[{\bf J}^{-1}({\bf q},{\bf f})]},
\end{equation}  
${\bf q}=[\tilde{x}_{1},\tilde{z}_{1},\tilde{g}_{1},...,\tilde{x}_{{\tilde{M}}},\tilde{z}_{{\tilde{M}}},\tilde{g}_{{\tilde{M}}}]$ is the 
vector of localization and power parameters out of the initial localization estimate $\tilde{\bf P}(\tilde{\bf f})$ and $\tilde{\bf G}$ defined in \eqref{eq:G}, 
${\bf J}({\bf q},{\bf f})$ is the Fisher information matrix with the $ij$th element \cite{crb}
\small
\begin{equation}
	[{\bf J}]_{ij}=T\, \mbox{tr}\left[ {\bf R}^{-1}_{\textrm{NFMC}}({\bf q},{\bf f})\frac{\partial{\bf R}_{\textrm{NFMC}}({\bf q},{\bf f})}{\partial q_i}
	{\bf R}^{-1}_{\textrm{NFMC}}({\bf q},{\bf f})\frac{\partial {\bf R}_{\textrm{NFMC}}({\bf q},{\bf f})}{\partial q_j}  \right],
\end{equation}  
\normalsize
where ${\bf R}_{\textrm{NFMC}}({\bf q},{\bf f})$ is defined in (34) with the corresponding notations, 
\small
\begin{equation}
	\frac{\partial{\bf R}_{\textrm{NFMC}}({\bf q},{\bf f})}{\partial q_i}\approx \frac{1}{2\Delta_i} \left[ {\bf R}_{\textrm{NFMC}}({\bf q}+\Delta_i {\bf e}_i,{\bf f})-
	 {\bf R}_{\textrm{NFMC}}({\bf q}-\Delta_i {\bf e}_i,{\bf f})\right],
\end{equation}  
\normalsize
where ${\bf e}_i$ is the vector of all zeros except $e_i=1$, and $\Delta_i>0$ is the control parameter for numerical derivative approximation as in \cite{crb}.\footnote{If we know the true model, it can be used for impedance optimization by employing ${\bf R}_\textrm{TM}$ instead of ${\bf R}_\text{NFMC}$, albeit it is computationally complex.} 

Formally, \eqref{eq:opt} is an unrestricted optimization problem of the non-convex multi-variable function. Any local optimization algorithm can be applied to obtain a locally optimized solution such that the conventional gradient or element-by-element iterative gradient searches for the given number of iterations as in \cite{em2} for 
the data rate maximization problem for similar EM modeling. Further simplification of the algorithms could be the introduction of some finite alphabet (FA) or even a binary search, especially considering that such search could be appropriate for for some particular implementations of tunable RIS elements, e.g., as in \cite{diode}. The RIS profile optimization procedure is summarized in Algorithm \ref{alg:ris_optimization}, 
where ${\cal F}$ is the optimization area, which may be a continuous area $\cal F=R$, or some FA set of $N_\text{FA}$ elements.

\begin{algorithm}[t!]
\caption{RIS Profile Optimization}
\label{alg:ris_optimization}
\begin{algorithmic}
\REQUIRE $\cal F$, ${\bf H}_{\textrm{NFMC}}({\bf q},{\bf f})$
\STATE \textbf{Initialization:} Random $\hat{f}_n \in {\cal F}$, $n=1,\ldots,N_{\text{R}}$
\REPEAT
    \STATE $\bar{\bf f} = \hat{\bf f}$
    \STATE $\bar{\nu} = \mbox{CRB}_{\textrm{NFMC}}({\bf q},\hat{\bf f})$
    \FOR{$n = 1,\ldots,N_{\text{R}}$}
        \STATE $\nu = \min_{\hat{f}_n \in {\cal F}} \mbox{CRB}_{\textrm{NFMC}}({\bf q},\hat{\bf f})$
		\STATE $ \tilde{f}_n=\text{arg }\min_{\hat{f}_n\in {\cal F}} \mbox{CRB}_{\textrm{NFMC}}({\bf q},\hat{\bf f})$
        \IF{$\bar{\nu} = \nu$}
            \STATE $\hat{f}_n = \bar{f}_n$
        \ELSE
            \STATE $\hat{f}_n=\tilde{f}_n$
            \STATE $\bar{\nu} = \nu$
        \ENDIF
    \ENDFOR
\UNTIL{$\hat{\bf f} = \bar{\bf f}$}
\ENSURE The optimized tunable RIS profile $\hat{\bf f}$
\end{algorithmic}
\end{algorithm}

\section{Simulation results}
We keep the main simulation assumptions as in Section V.A if different is not specified.  
For signal based EL mismatch analysis and direct localization with no RIS, 
we assume LOS propagation conditions modeled by equations \eqref{eq:HEMTr}, \eqref{eq:PSI0}, \eqref{eq:PSIrrdir}, \eqref{eq:PSIrtdir}, 
and simulate a uniform planar array with $N=64$ elements. Also, we consider $T=10$ data samples, $h=0.5$ m, and $\sigma^2=-87$ dBm. 

\subsection{Signal based EL model mismatch analysis}
The mismatch metric $\hat{\gamma}_{\textrm{MM}}({\bf p}_s)$ in \eqref{eq:MMsig}, i.e., the ratio between the empirical average of LR for mismatched model and the mean of pre-known LR considering the true model, is plotted in Figs. 6 and 7 for various location of the source, considering FF and NF models without and with MC correction, respectively. The LR is averaged over $K=50$ independent scenario realizations for different transmit powers, $g \in \{0,10,20\}$ dBm, using the analytical EM model as the TM for signal generation. Thus, values of $\hat{\gamma}_{\textrm{MM}}({\bf p}_s)$ close to one suggest that the mismatched model exhibits statistically similar to the true model. Conversely, smaller values of $\hat{\gamma}_{\textrm{MM}}({\bf p}_s)$ indicate a significant mismatch relative to the true model.

\begin{figure}[!t]
\centering
\includegraphics[width=80mm,height=80mm]{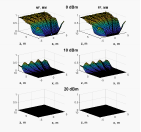} 
\caption{Mismatch metric (30) for NF and FF MMs without MC correction, for various transmit powers of $0$, $10$, and $20$ dBm.}
\label{fig:LRnoMC}
\end{figure}
\begin{figure}[!htb]
\centering
\includegraphics[width=80mm,height=80mm]{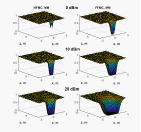} 
\caption{Mismatch metric (30) for NFMC and FFMC MMs with MC correction, for various transmit powers of $0$, $10$, and $20$ dBm.}
\label{fig:LR}
\end{figure}

\begin{figure}[!t]
\centering
\includegraphics[width=0.95\linewidth,height=85mm]{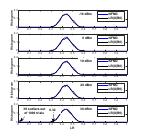} 
\vspace{-0.5cm}
\caption{${\mbox{LR}}$ estimates for NFMC with $M=1$,  ${\bf  p}=[-2,-0.5,4]^{\text{T}}$m,\\ and $g\in \{-10,0,10,20,30\}$ dBm.}
\label{fig:LRNFMC}
\end{figure}

To evaluate the performance of localization using the mismatched model NFMC, we present in Fig.~\ref{fig:LRNFMC} the histogram of the ${\mbox{LR}}$ metric when NFMC is employed for positioning via Algorithm~\ref{alg:model_location_classification}. The corresponding MM is included as a reference to the ML algorithm defined in \eqref{eq:PMM}, where NFMC serves as the mismatched model. The histogram is compared against the model-independent p.d.f. of the ${\mbox{LR}}$ for the true model. We consider the case of $M=1$ with ${\bf p} = [-2, -0.5, 4]^{\text{T}}$~m for various transmit power levels. The MUSIC initialization is defined over a $15 \times 15$ uniform grid within an area of $1.4 \times 1.4$~m centered at the source position. Optimization for \eqref{eq:PMM} is performed using MATLAB's ``fminsearch'' routine. The results presented are averaged over 1000 scenario realizations. 

In Fig.~\ref{fig:LRNFMC}, the NFMC model demonstrates negligible model mismatch for transmit power levels of $g = [-10, 0, 10, 20]$~dBm, with no apparent outliers and histograms that closely align with the predefined distribution. However, this consistency breaks for the highest transmit power level, $g = 30$~dBm, as shown in the last subfigure of Fig.~\ref{fig:LRNFMC}, where a mismatch is observed. To detect outliers, we compute the threshold $\beta(64,10) = 0.32$ for $p_{\beta} = 0.01$ as specified in \eqref{eq:beta1} and \eqref{eq:beta2}, using Algorithm~\ref{alg:model_location_classification}. Under these settings, $39$ out of 1000 trials are identified as outliers for $g = 30$~dBm, as shown in Fig.~\ref{fig:LRNFMC}.

\begin{figure}[!t]
\centering
\includegraphics[width=0.95\linewidth]{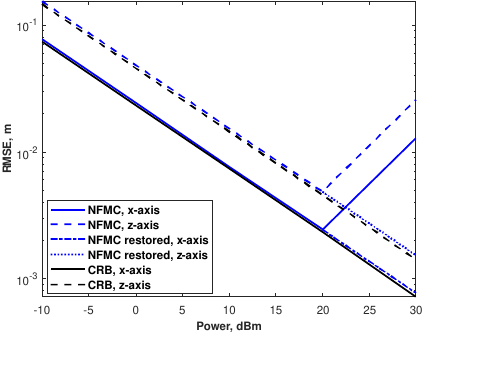} 
\caption{RMSE estimates for NFMC with $M=1$, ${\bf  p}=[-2,-0.5,4]^{\text{T}}$m,\\ and $g\in \{-10,0,10,20,30\}$ dBm.}
\label{fig:RMSENFMC}
\end{figure}

The conclusions derived from the ${\mbox{LR}}$ metric in Fig.~\ref{fig:LRNFMC} are further validated in Fig.~\ref{fig:RMSENFMC} through the root-mean-square error (RMSE) for positioning accuracy under the same settings. The CRLB for the true model in this scenario is estimated as outlined in Section~\ref{subsec:tunableRIS} with $\Delta = 0.001$~m and is included in Fig.~\ref{fig:LRNFMC} as a benchmark for both the $x$ and $y$ components of the source position. For $g \leq 20$~dBm, the localization performance of the mismatched model NFMC matches the CRLB for the true model, indicating no significant model mismatch. Conversely, for $g > 20$~dBm, the RMSE deviates from the CRLB and increases with rising transmit power, signifying a notable mismatch. 

This degraded performance can be attributed to the outliers detected in Fig.~\ref{fig:LRNFMC}, which may result from poor MUSIC initialization in this scenario. The strength of our algorithm lies in its ability to detect and handle outliers. By updating the MUSIC initialization for the identified $39$ outliers, such as performing spectral searches over a finer $50 \times 50$ grid instead of the $15 \times 15$ grid used previously, the RMSE can be restored to align with the corresponding true model CLRB. This improvement is reflected in Fig.~\ref{fig:LRNFMC} as the restored NFMC.

\begin{figure}[!t]
\centering
\includegraphics[width=95mm,height=60mm]{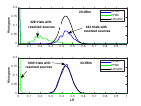} 
\caption{${\mbox{LR}}$ estimates for $M=2$,  ${\bf  p}_1=[-2,-0.5,4]^{\text{T}}$m, ${\bf  p}_{2}=[-2.4,-0.5,4.4]^{\text{T}}$m, and $g=[20,30]$ dBm.}
\label{fig:LRtwosources}
\end{figure}

\begin{table}[th!]
\centering
\caption{TM CRB and RMSE estimates for $M=2$, ${\bf p}_{1} = [-2, -0.5, 4]^{\text{T}}$~\textnormal{m}, ${\bf p}_{2} = [-2.4, -0.5, 4.4]^{\text{T}}$~\textnormal{m}, and $g = [20, 30]$~\textnormal{dBm} for trials with resolved sources.}
\label{Table:RMSEtwosources}
\renewcommand{\arraystretch}{1.3} 
\setlength{\tabcolsep}{8pt} 
\begin{tabular}{|>{\columncolor[gray]{0.9}}p{1cm}|c|c|>{\columncolor[gray]{0.95}}c|>{\columncolor[gray]{0.95}}c|>{\columncolor[gray]{0.9}}c|}
\hline
\textbf{Power (dBm)} & \textbf{Axis} & \textbf{Source} & \multicolumn{2}{c|}{\cellcolor[gray]{0.8}\textbf{RMSE (m)}} & \textbf{CRB (m)} \\ \cline{4-5}
                     &               &                 & \cellcolor[gray]{0.85}\textbf{NFMC} & \cellcolor[gray]{0.85}\textbf{FFMC} &                  \\ \hline
\rowcolor[gray]{0.95} 20             & x             & 1               & 0.0112             & 0.0147            & 0.0108           \\ \cline{3-6}
\rowcolor[gray]{1}                   &               & 2               & 0.0156             & 0.0300            & 0.0150           \\ \cline{2-6}
\rowcolor[gray]{0.95}                & z             & 1               & 0.0121             & 0.0151            & 0.0115           \\ \cline{3-6}
\rowcolor[gray]{0.97}                   &               & 2               & 0.0161             & 0.0316            & 0.0156           \\ \hline
\rowcolor[gray]{0.95} 30             & x             & 1               & 0.0038             & 0.0142            & 0.0036           \\ \cline{3-6}
\rowcolor[gray]{0.97}                   &               & 2               & 0.0053             & 0.0284            & 0.0051           \\ \cline{2-6}
\rowcolor[gray]{0.95}                & z             & 1               & 0.0040             & 0.0137            & 0.0038           \\ \cline{3-6}
\rowcolor[gray]{0.97}                   &               & 2               & 0.0056             & 0.0298            & 0.0053           \\ \hline
\end{tabular}
\end{table}
%
The ${\mbox{LR}}$ and RMSE results for $M=2$, with ${\bf p}_{1} = [-2, -0.5, 4]^{\text{T}}$~m and ${\bf p}_{2} = [-2.4, -0.5, 4.4]^{\text{T}}$~m, are shown in Fig.~\ref{fig:LRtwosources} and Table~\ref{Table:RMSEtwosources}, respectively. These results were obtained using the ML approach with precise MUSIC initialization based on a $50 \times 50$ grid to mitigate initialization issues. For $g = 20$~dBm, Fig.~\ref{fig:LRtwosources} reveals the reduced resolution capability of FFMC compared to NFMC. This limitation is also reflected in Table~\ref{Table:RMSEtwosources}, where the positioning accuracy of FFMC is notably lower than the TM CRB. At $g = 30$~dBm, both sources are resolved in all trials for NFMC and FFMC, as shown in Fig.~\ref{fig:LRtwosources}. However, the FFMC estimates are classified as outliers in all trials due to ${\mbox{LR}}[{\bf X}|\hat{\bf R}_{\text{FFMC}}(\hat{\bf P}^{\text{FFMC}})] \ll \beta(64,10)$. In contrast, NFMC significantly outperforms FFMC in terms of RMSE, as demonstrated in Table~\ref{Table:RMSEtwosources}. This highlights a substantial FFMC model mismatch, even when the distance to both sources exceeds the Fraunhofer distance of 0.7~m.

The comparison of RMSE with the CRB for the known TM underscores the presence of a model mismatch in this scenario. The unique advantage of EL-based mismatch analysis lies in its ability to classify a model as true or mismatched without requiring prior knowledge of the TM or a ``good" MM. Moreover, this classification can be performed on a realization-by-realization basis, providing a flexible and powerful tool for model evaluation.

\subsection{Non-uniform RIS assisted localization}
\begin{figure}[!t]
\centering
\includegraphics[width=80mm,height=55mm]{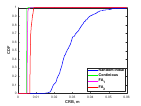} 
\caption{Steady state CRB for Algorithm 2 for different optimization areas for ${\bf p}=[-1.51,-1,6.61]^{\text{T}}$ m.}
\label{fig:opt_ris_1}
\end{figure}
\begin{figure}[!t]
\centering
\includegraphics[width=80mm,height=55mm]{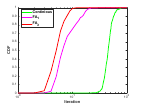} 
\caption{Convergence rate for Algorithm 2 for different optimization areas for ${\bf p}=[-1.51,-1,6.61]^{\text{T}}$ m.}
\label{fig:opt_rate_1}
\end{figure}

For RIS-related simulations, we use the non-uniform BS array and RIS configurations specified in Section V.A for the EM propagation channels modeled by Equations \eqref{eq:HEMTr} - \eqref{eq:PSIrr}, \eqref{eq:PSIrtRIS} for $h=1$ m, and $\sigma^2 = -120$ dBm.

We begin by illustrating the RIS profile optimization developed in Section V.B. We simulate Algorithm~\ref{alg:ris_optimization} for NFMC in the single-source scenario, with $M=1$, ${\bf p}=[-1.51, -1, 6.61]^{\text{T}}$~m, and $g = 0$~dBm. The optimization is performed over the following areas: \textit{(i)} continuous area ${\cal F}=[-\xi,\xi]$; \textit{(ii)} finite alphabet area $1$, $\mbox{FA}_1=\mu [-2,-1,0,1,2]$; and \textit{(iii)} finite alphabet area $2$, $\mbox{FA}_2=\mu [-1,1]$, where $\mu = 100$. We generate a random profile as ${\bf f} \sim {\cal{RN}}\{{\bf 0}, \mu {\bf I}_{N_{\text{R}}}\}$. For continuous optimization over ${\cal F}=[-\xi, \xi]$, we use the ``fminbnd" MATLAB routine with $\xi = 500$.

The steady-state CRB and convergence rate results for Algorithm~\ref{alg:ris_optimization} across different optimization areas are shown in Figs. \ref{fig:opt_ris_1} and \ref{fig:opt_rate_1}, based on 500 trials. One iteration in Fig. \ref{fig:opt_rate_1} corresponds to the sequential adjustment of all $N_{\text{R}}$ elements of the vector ${\bf f}$. From these figures, the following observations can be made: the optimization over the continuous ${\cal F}$ yields the best steady-state CRB results, as shown in Fig. \ref{fig:opt_ris_1}, though it comes at the cost of very slow convergence, as illustrated in Fig. \ref{fig:opt_rate_1}. The steady-state CRB performance for $\mbox{FA}_1$ is nearly identical to the continuous ${\cal F}$ optimization, with only slight performance degradation observed for the simplest binary $\mbox{FA}_2$ optimization area, as seen in Fig. \ref{fig:opt_ris_1}. Finally, the simplest binary version of Algorithm~\ref{alg:ris_optimization} presents a promising RIS optimization approach in terms of complexity and convergence rate in the considered scenario.

%
\begin{figure}[!t]
\centering
\includegraphics[width=80mm,height=55mm]{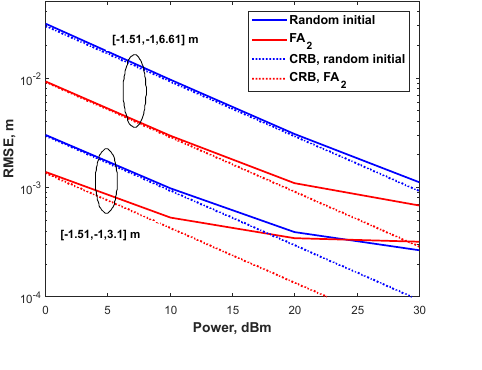} 
\caption{Average RMSE and CRB performance for ${\cal F}=\mu [-1,1]$ and $\gamma=1.01$ for different source locations.}
\label{fig:rmse_ris_2}
\end{figure}
\begin{figure}[!t]
\centering
\includegraphics[width=80 mm,height=70mm]{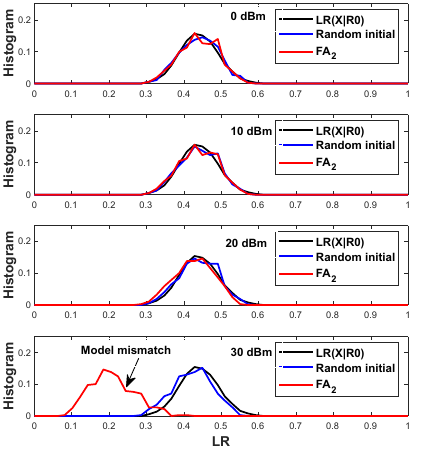} 
\caption{LR distributions in the scenario in Fig. \ref{fig:rmse_ris_2} for  ${\bf p}=[-1.51,-1,6.61]^{\text{T}}$ m.}
\label{fig:lr_ris_5}
\end{figure}
\begin{figure}[!t]
\centering
\includegraphics[width=80 mm,height=70mm]{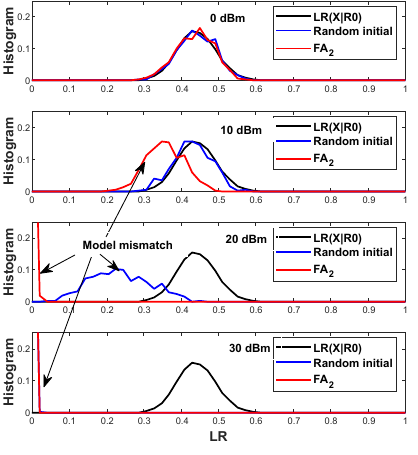} 
\caption{LR distributions in the scenario in Fig. \ref{fig:rmse_ris_2} for  ${\bf p}=[-1.51,-1,3.1]^{\text{T}}$ m.}
\label{fig:lr_ris_4}
\end{figure}
The average RMSE and CRB results over 500 trials are shown in Fig.~\ref{fig:rmse_ris_2} for the binary version of Algorithm~2 within the RIS optimization protocol, as presented in Fig.~\ref{fig:ris_opt_1}. These results are obtained in the single-source scenario for different positions, ${\bf p}=[-1.51,-1,6.61]^{\text{T}}$ m and ${\bf p}=[-1.51,-1,3.1]^{\text{T}}$ m, with $T_1=T_2=10$. The corresponding LR results are presented in Figs.~\ref{fig:lr_ris_5} and \ref{fig:lr_ris_4}. For the ML initialization in \eqref{eq:PMM} for the NFMC localization model, we use the MUSIC algorithm \eqref{eq:music} computed over a $100 \times 100$ grid in the area $x=[-3,1]$ m and $z=[1,8]$ m.

From Figs.~\ref{fig:rmse_ris_2}, \ref{fig:lr_ris_5}, and \ref{fig:lr_ris_4}, the following observations can be made. The RMSE performance in Fig.~\ref{fig:rmse_ris_2} is very close to the CRB for both the initial and optimized RIS profiles in scenarios with no NFMC model mismatch. In these cases, the corresponding LR values align with the predefined LR distribution, as shown in Fig.~\ref{fig:lr_ris_5} for a signal power of [0, 10, 20] dBm and ${\bf p}=[-1.51,-1,6.61]^{\text{T}}$ m, and in Fig.~\ref{fig:lr_ris_4} for 0 dBm signal power with ${\bf p}=[-1.51,-1,3.1]^{\text{T}}$ m. However, for higher signal power, a significant NFMC model mismatch is observed, particularly in Figs.~\ref{fig:lr_ris_5} and \ref{fig:lr_ris_4}. The mismatch is especially pronounced in Fig.~\ref{fig:lr_ris_4} for the source ${\bf p}=[-1.51,-1,3.1]^{\text{T}}$ m, which is closer to the array and RIS. This mismatch leads to notable performance degradation compared to the potential CRB accuracy in Fig.~\ref{fig:rmse_ris_2}.

It is important to emphasize that the identification of localization model match/mismatch areas cannot be made through RMSE/CRB analysis, as shown in Fig.~\ref{fig:rmse_ris_2}, unless the true model is known. In contrast, the proposed EL approach enables a realization-by-realization model mismatch analysis, as demonstrated in Figs.~\ref{fig:lr_ris_5} and \ref{fig:lr_ris_4} for the unknown TM.
\begin{figure}[!t]
\centering
\includegraphics[width=80mm,height=55mm]{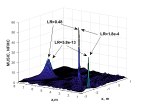} 
\caption{Example of NFMC MUSIC pattern for $M=\hat{M}=2$ for ${\bf p}_{1}=[-1.51,-1,6.61]^{\text{T}}$ m, ${\bf p}_{2}=[-1.51,-1,3.1]^{\text{T}}$ m, $g_1=g_2=10$ dBm.}
\label{fig:m2_3_example_1}
\end{figure}

Although the use of the non-uniform minimal redundancy array and RIS configurations improves the localization area, it does not guarantee the absence of false MUSIC peaks in certain scenarios, particularly when using specific RIS profiles in multi-source situations. An example of such a scenario is shown in Fig.~\ref{fig:m2_3_example_1} for $M=2$, where ${\bf p}_1=[-1.51,-1,6.61]^{\text{T}}$ m, ${\bf p}_2=[-1.51,-1,3.1]^{\text{T}}$ m, and a random RIS profile are used in the same network configuration as depicted in Figs.~\ref{fig:opt_ris_1} to \ref{fig:lr_ris_4}. In this case, the MDL detection estimates $\hat{M}=2$, but it is evident from Fig.~\ref{fig:m2_3_example_1} that there are $3$ distinct peaks in the MUSIC pattern.

The EL analysis resolves this issue by grouping the corresponding estimates into $\hat{M}=2$ sources and calculating the corresponding LR values using \eqref{eq:LRMM}, as shown in Fig.~\ref{fig:m2_3_example_1}. This correct grouping results in an LR value of $0.48$, which aligns with the predefined LR distribution in Fig.~\ref{fig:lr_distr_3} for $N=64$ and $T=10$. Other groupings lead to very low LR values, indicating incorrect estimates. Clearly, this analysis can be conducted in scenarios without significant localization model mismatch, where any incorrect grouping would yield low LR values outside the predefined LR distribution. In such cases, localization model correction might be necessary.


\section{Conclusions}
We propose a framework exploiting the expected likelihood metric to analyze model mismatches for localization purposes along with an online (on-the-fly) positioning algorithm with model selection and outlier detection. In this framework, we explored using the analytical electromagnetic model as the true model, while considering near-field  and far-field models with mutual coupling corrections as mismatched models. A key advantage of the EL approach over misspecified CRLB-based investigations is its ability to perform online, realization-by-realization model mismatch analysis without requiring knowledge of the true model. Instead, it compares the likelihood ratio of the given model and localization estimates with a predefined, scenario-independent LR distribution derived from the actual covariance matrix. The applicability and efficiency of the EL analysis have been demonstrated through simulations utilizing the analytical electromagnetic model for signal generation. These simulations consider various simplified models for source localization, both in direct and reconfigurable intelligent surface (RIS)-assisted localization scenarios. Significant model mismatches are observed when the signal power increases, leading to noticeable deviations from the true model’s behavior. The LR analysis effectively detects these mismatches, with the LR framework consistently corresponding to the RMSE conclusions. This demonstrates that the LR analysis provides a robust and reliable tool for model mismatch detection and correction in real-world localization scenarios.


\bibliographystyle{IEEEtran}

\bibliography{IEEEtranBSTCTL,ElzanatyBibtex}
\end{document}